\documentclass[12pt]{amsart}
\newcommand{\R}{\mathbf R}
\newcommand{\Z}{\mathbf Z}

\newcommand{\DD}{\mathcal D}
\newcommand{\CC}{\mathcal C}

\newcommand{\LL}{\mathcal L}
\newcommand{\BB}{\mathcal B}
\newcommand{\HH}{\mathcal H}

\newcommand{\Vect}{Vect}

\newcommand{\g}{\mathfrak g}

\newtheorem{theorem}{Theorem}
\newtheorem{lemma}{Lemma}
\numberwithin{equation}{section}
\numberwithin{theorem}{section}
\numberwithin{lemma}{section}

\begin{document}
\title{Quantization of multidimensional variational principles}
\author{Alexander Roi Stoyanovsky}
\begin{abstract}
We construct a mathematical version of quantum field theory. It assigns to a multidimensional variational principle an associative algebra which is a quantization 
of the Poisson algebra of classical field theory observables. For free scalar field and for the Dirichlet principle, this construction yields free algebraic quantum field theory 
and free field representation in two dimensional conformal field theory.  
\end{abstract}
\maketitle

\section*{Introduction}

The main question addressed in this paper is the following: what is mathematics behind quantum field theory (QFT)?

A known example of such mathematics is two dimensional conformal field theory, see e.~g. [\ref{BPZ}, \ref{FF}, \ref{Fr}]. This theory yields some ``strange'' formulas with normally ordered products and 
operator product expansions, arising in representaton theory of infinite dimensional Lie algebras and in related algebraic geometry. The present research originated from attempts to mathematically understand
these formulas. The result is a mathematical version of QFT presented in this paper.

Traditional perturbative and axiomatic QFT provides a computational and conceptual framework for the physical theory of elementary particles and their interactions. However, mathematical realization of this theory
met serious difficulties. Up to now mathematical realization of QFT axioms for realistic theories in 4-dimensional spacetime has not been obtained.

 In fact, the reason for these difficulties seems to be related with deep logical problems of QFT. Indeed, this theory uses the notions of vacuum and particles.
However [\ref{BS}], the particles always interact with vacuum as with the medium in which they move. In ``right'' QFT there should be no natural notions of medium or particles, but there should be only their synthesis, 
the universal form of matter called the interacting quantum field. 

A way to search for ``right'' QFT is given by the Dirac principle: one should choose an appropriate mathematical theory and develop it in appropriate direction, however unexpected would be the obtained physical picture.

As such a mathematical theory, we choose the theory of linear partial differential equations (PDEs). It is a rich beautiful theory, which played a central role in XX century mathematics. It yields a mathematical 
formulation for wave theory, generalizing both quantum mechanics and wave optics.

The idea of Feynman path integral can mean that mathematics of QFT, if it exists, should be a generalization of the theory of linear PDEs
to the case of multidimensional variational principles and multidimensional bicharacteristics. Thus, our main question is: how to generalize the theory of linear PDEs to the case of multidimensional variational principles?

Following this way, in a many year research [\ref{St07}, etc.] the author has defined and studied generalizations of the Hamilton--Jacobi and Schr\"odinger equations for multidimensional variational principles. The
generalized Hamilton--Jacobi equation is a well defined nonlinear first order functional differential equation (i.~e. equation for a functional involving its first order functional derivatives), 
whose integration theory has been shown to amount to the integration theory of 
the Euler--Lagrange equation or the equivalent generalized canonical Hamilton equations. However, the generalized Schr\"odinger equation has been stated only 
formally. For a long time the author tried to find a mathematical interpretation of this equation and its relation with QFT.
The main difficulty is that one cannot define the notion of solution of this equation. This problem is closely related with the problem of defining 
dynamical evolution in QFT, because the Schr\"odinger equation is usually considered as an equation for dynamical evolution of a system in time. 
It turns out [\ref{MS}, \ref{St07}] that dynamical evolution in QFT is always well defined only in the principal order of quasiclassical approximation, 
i.~e. in the one-loop approximation in perturbative QFT. In general, as a rule, dynamical evolution does not exist.

In the present paper we propose a way out of these difficulties. The idea is very simple. Since in general we cannot define solutions (states), let us try to 
define observables. They should form an associative algebra. 
It turns out that we can define the algebra of observables directly from the generalized Schr\"odinger equation. 
This algebra is a quantization of the commutative Poisson algebra of classical field theory observables. This construction is the main result of the paper. The key point in it 
is that we consider the spacetime variables and the field variables on equal footing. That is, we consider the usual Schr\"odinger equation not as an evolutionary equation but as an equation in which
the time variable and the coordinate variables are considered mathematically on equal footing. We assign to this equation the algebra of quantum mechanical observables. 
Generalizing this construction to multidimensional case, we obtain the required construction of the algebra of QFT observables.

The paper is organized as follows. 
\nopagebreak

\S1 is devoted to the formalism of classical field theory. We consider classical field theories given by variational principles for first order 
action functionals with boundary terms, invariant with respect to certain symmetry group.
Such variational principles include
the Einstein general relativity. For them one has a generalization of the Noether theorem on symmetries, see \S\ref{subsect_Noether}.
Classical field theory assigns to such an invariant variational principle an equivariant commutative Poisson algebra of classical observables (functionals
of a solution of the Euler--Lagrange equation).

In \S2 we construct an equivariant associative algebra which is a quantization of the equivariant Poisson algebra of classical observables, for a wide class of field theories including physically and mathematically important cases.

In \S3 we show that for free scalar field in the Minkowski spacetime and for the Dirichlet principle, our quantization scheme yields the standard free algebraic QFT [\ref{TV}] and the free field representation in two-dimensional 
conformal field theory [\ref{BPZ}, \ref{FF}, \ref{Fr}].

Finally, in Conclusion we discuss relations of our theory with standard QFT.

\section{Classical field theory for variational principles with boundary terms}
\label{classical}

\subsection{Variational principles with boundary terms} We shall consider 
classical field theories given by variational principles for the invariant integral, called action, of the form
\begin{equation}
\label{action}
J=\int_\DD L(x,\varphi(x),\varphi_x(x))\,dx+\int_\CC B(x(s),\varphi(x(s)),\varphi_x(x(s)),x_s(s))\,ds,
\end{equation} 
where $x=(x^0,\ldots,x^n)$ is a spacetime point; 
$\DD$ is a domain in the spacetime $\R^{n+1}$ with the smooth boundary $\CC=\partial\DD$;
$\varphi(x)=(\varphi^i(x))$ are real smooth field
functions,  $i=1,\ldots,M$; $\varphi_x=(\varphi^i_{x^j})=(\partial\varphi^i/\partial x^j)$, $j=0,\ldots,n$; 
$s=(s^1,\ldots,s^n)$ are parameters on the boundary surface $\CC: x=x(s)$; and $x_s=(x^j_{s^k})=(\partial x^j/\partial s^k)$, $k=1,\ldots,n$. 
The function $L$ is called the Lagrangian, and 
$B$ is called the boundary term. We require that $B$ is a density in $s$ of degree 1. This means that for any 
$n\times n$-matrix $(s^{k'}_{t^k})_{1\le k,k'\le n}$ with
$\det(s^{k'}_{t^k})>0$,
we have 
\begin{equation}
\label{B_density_in_s}
B\left(x,\varphi,\varphi_x,\sum_{k'=1}^n x^j_{s^{k'}}s^{k'}_{t^k}\right)=\det(s^{k'}_{t^k}) B(x,\varphi,\varphi_x,x^j_{s^k}).
\end{equation}
This formula implies that 
the boundary integral in (\ref{action}) is independent on the choice of parameters $s$ on the (oriented) boundary surface $\CC$. 

Invariance of action means that integral (\ref{action}) is preserved by certain Lie group $G$ of diffeomorphisms of the space $\R^{n+1+M}$ with coordinates $(x^j,\varphi^i)$. The group $G$ is called the {\it symmetry group} of the theory.

{\it Examples.} 1) The scalar self-interacting field. Let $M=1$ and
\begin{equation}
\label{scalar_field}
L(x,\varphi,\varphi_x)=\frac 12\left(\varphi_{x^0}^2-\sum_{j=1}^n\varphi_{x^j}^2\right)-V(\varphi),\ \ B=0,
\end{equation}
where $V(\varphi)$ is any function. The symmetry group is the Poincar\'e group.

2) The Dirichlet principle. Let $M=1$, $n=1$, and 
\begin{equation}
\label{Dirichlet}
L(x,\varphi,\varphi_x)=\frac12\left(\varphi_{x^0}^2+\varphi_{x^1}^2\right),\ \ B=0.
\end{equation} 
The symmetry group is the group of conformal transformations of the plane.

3) The Einstein general relativity (see e.~g. [\ref{D}]). The symmetry group is the group of diffeomorphisms of spacetime. Note that in this example the boundary term is nonzero.

4) The Yang--Mills theory (see e.~g. [\ref{IZ}]). The symmetry group is the semidirect product of the Poincar\'e group and the gauge group.

The main idea of our approach is to consider the variables $x^j$ and $\varphi^i$ on equal rights. This is achieved by rewriting integral (\ref{action}) in the parametric form. 
A variational principle in the parametric form is given by the invariant integral, called parametric action,  
\begin{equation}
\label{parametric_action}
J=\int_\DD\LL(y(u),y_u(u))\,du+\int_\CC\BB(y(u(s)),y_u(u(s)),u_s(s))\,ds,
\end{equation}
where $u=(u^0,\ldots,u^n)$; $\DD$ is a domain in $\R^{n+1}$ with the smooth boundary $\CC=\partial\DD$;
$y(u)=(y^l(u))$ are real smooth functions, $l=1,\ldots,N$; $y_u=(y^l_{u^j})=(\partial y^l/\partial u^j)$, $j=0,\ldots,n$;
$s=(s^1,\ldots,s^n)$ are parameters on the boundary surface $\CC:u=u(s)$;
and $u_s=(u^j_{s^k})=(\partial u^j/\partial s^k)$, $k=1,\ldots,n$. We require that the boundary term $\BB$ is a density in $s$ of degree 1, cf. (\ref{B_density_in_s}). This 
implies that the boundary integral in (\ref{parametric_action}) is independent on the choice of parameters $s$ on the (oriented) boundary surface $\CC$. We also require that 
the Lagrangian $\LL$ is a density in $u$ of degree 1, and the boundary term $\BB$
is a density in $u$ of degree~0. This means that
for any $(n+1)\times(n+1)$-matrix $(u^{j'}_{w^j})_{0\le j,j'\le n}$ with
$\det(u^{j'}_{w^j})>0$,
we have
\begin{equation}
\label{density_LL_BB}
\begin{aligned}
&\LL\left(y,\sum_{j'=0}^n y^l_{u^{j'}}u^{j'}_{w^j}\right)=\det(u^{j'}_{w^j})\LL(y,y^l_{u^j}),\\ 
&\BB\left(y,\sum_{j'=0}^n y^l_{u^{j'}}u^{j'}_{w^j},\sum_{j'=0}^n w^j_{u^{j'}}u^{j'}_{s^k}\right)=\BB(y,y^l_{u^j},u^j_{s^k}),
\end{aligned}
\end{equation}
where $(w^j_{u^{j'}})$ is the inverse matrix to the matrix $(u^{j'}_{w^j})$.
This formula implies that the whole integral (\ref{parametric_action}) is preserved by arbitrary (orientation preserving) changes of parameters $u$. 
Invariance of parametric action (\ref{parametric_action}) means that it is preserved by certain Lie group $G$ of diffeomorphisms of the space $\R^N$ with coordinates $y$. The group
$G$ is called the symmetry group of the theory. 

{\it Example.} The Plateau problem: the area of a surface in the Euclidean space. Let 
\begin{equation}
\label{Plateau}
\LL(y,y_u)=\sqrt{\sum_{1\le l_0<\ldots<l_n\le N}\left(\frac{\partial(y^{l_0},\ldots,y^{l_n})}{\partial(u^0,\ldots,u^n)}\right)^2},\ \  \BB=0,
\end{equation}
where 
\begin{equation}
\label{Jac}
\frac{\partial(y^{l_0},\ldots,y^{l_n})}{\partial(u^0,\ldots,u^n)}=\det(y^{l_j}_{u^{j'}})_{0\le j,j'\le n}
\end{equation}
is the Jacobian. Then parametric action (\ref{parametric_action}) equals the area of the $(n+1)$-dimensional surface $(y(u),u\in\DD)$ in the Euclidean space $\R^N$.
The symmetry group is the group of affine orthogonal transformations of the space $\R^N$.

Obviously, parametric action (\ref{parametric_action}) is a particular case of action (\ref{action}) with 
$M=N$, $x=u$, $\varphi=y$, $L=\LL$, and $B=\BB$. 
The symmetry group of this action (\ref{action}) is the direct product of the symmetry group of parametric action (\ref{parametric_action}) and the group of (orientation preserving) diffeomorphisms of the variables $x$. 
Conversely, any action $(\ref{action})$ can be written in the parametric form (\ref{parametric_action}) with $N=n+1+M$, $y=(x,\varphi)$, the same symmetry group $G$, and
\begin{equation}
\label{action_to_param_action}
\begin{aligned}
&\LL(y,y_u)=\LL(x,\varphi,x_u,\varphi_u)=L\left(x,\varphi,\sum_{j'=0}^n\varphi^i_{u^{j'}}u^{j'}_{x^j}\right)\det(x^j_{u^{j'}})_{0\le j,j'\le n},\\
&\BB(y,y_u,u_s)=\BB(x,\varphi,x_u,\varphi_u,u_s)=B\left(x,\varphi,\sum_{j'=0}^n\varphi^i_{u^{j'}}u^{j'}_{x^j},\sum_{j'=0}^n x^j_{u^{j'}}u^{j'}_{s^k}\right),
\end{aligned}
\end{equation}
where $(u^{j'}_{x^j})$ is the inverse matrix to the matrix $(x^j_{u^{j'}})$.

\subsection{Formula for variation of action}
\nopagebreak
\subsubsection{Parametric case} Let $y^l(u,\varepsilon)$ be a smooth function of $u\in\DD$ and of a real variable $\varepsilon$, which is a deformation of the function 
$y^l(u)=y^l(u,0)$, $1\le l\le N$. Denote by $\delta$ the variation, i.~e. the differential with respect to $\varepsilon$ at $\varepsilon=0$. Then,
integrating by parts, we have the following formula for variation of parametric action (\ref{parametric_action}):
\begin{equation}
\label{variation_of_parametric_action}
\begin{aligned}
\delta J&=\int_\DD\sum_{l=1}^N\left(\LL_{y^l}-\sum_{j=0}^n\frac\partial{\partial u^j}\LL_{y^l_{u^j}}\right)\delta y^l(u)du \\
&+\int_\CC\left(\sum_{l=1}^N v_l(s)\delta y^l(u(s))+\delta\BB\right)ds,
\end{aligned}
\end{equation}  
where $\LL_{y^l}=\partial\LL/\partial{y^l}$, $\LL_{y^l_{u^j}}=\partial\LL/\partial y^l_{u^j}$, and
\begin{equation}
\label{v_l}
v_l=\sum_{j=0}^n \LL_{y^l_{u^j}}(-1)^j\frac{\partial(u^0,\ldots,\widehat{u^j},\ldots,u^n)}{\partial(s^1,\ldots,s^n)};
\end{equation}
here $\partial(\ldots)/\partial(\ldots)$ is the Jacobian, cf.~(\ref{Jac}), and the hat over the variable $u^j$ in the Jacobian means that the variable is omitted.

The stationarity condition $\delta J=0$ for any $\delta y^l(u)$ with $\delta y^l(u(s))=\delta y^l_{u^j}(u(s))\equiv0$ yields the Euler--Lagrange equations
\begin{equation}
\label{param_Euler_Lagrange}
\LL_{y^l}-\sum_{j=0}^n\frac\partial{\partial u^j}\LL_{y^l_{u^j}}=0,\ \ l=1,\ldots,N.
\end{equation}

\subsubsection{Non-parametric case} Substituting (\ref{action_to_param_action}) into (\ref{variation_of_parametric_action}), we obtain the formula for variation of action (\ref{action}),
\begin{equation}
\label{variation_of_action}
\begin{aligned}{}
\delta J&=\int_\DD\sum_{i=1}^M\left(L_{\varphi^i}-\sum_{j=0}^n\frac{\partial}{\partial x^j}L_{\varphi^i_{x^j}}\right)\delta\varphi^i(x)dx\\
&+\int_\CC\left(\sum_{i=1}^M\pi_i(s)\delta\varphi^i(s)-\sum_{j=0}^n H_j(s)\delta x^j(s)+\delta B\right)ds,
\end{aligned}
\end{equation}
where 
\begin{equation}
\label{pi,H}
\begin{aligned}{}
\pi_i&=\sum_{j=0}^n L_{\varphi^i_{x^j}}(-1)^j\frac{\partial(x^0,\ldots,\widehat{x^j},\ldots,x^n)}{\partial(s^1,\ldots,s^n)},\\
H_j&=\sum_{j'\ne j}\left(\sum_{i=1}^M L_{\varphi^i_{x^{j'}}}\varphi^i_{x^j}\right)(-1)^{j'}\frac{\partial(x^0,\ldots,\widehat{x^{j'}},\ldots,x^n)}{\partial(s^1,\ldots,s^n)}\\
&+\left(\sum_{i=1}^M L_{\varphi^i_{x^j}}\varphi^i_{x^j}-L\right)(-1)^j\frac{\partial(x^0,\ldots,\widehat{x^j},\ldots,x^n)}{\partial(s^1,\ldots,s^n)}.
\end{aligned}
\end{equation}

The stationarity condition $\delta J=0$ for any $\delta\varphi^i(x)$ with $\delta\varphi^i(x(s))=\delta\varphi^i_{x^j}(x(s))\equiv0$ and for $\delta\varphi^i(s)=\delta x^j(s)\equiv0$ yields the Euler--Lagrange equations
\begin{equation}
\label{Euler_Lagrange}
L_{\varphi^i}-\sum_{j=0}^n\frac{\partial}{\partial x^j}L_{\varphi^i_{x^j}}=0,\ \ i=1,\ldots,M.
\end{equation}

\subsection{Hamiltonian formalism}

\subsubsection{Non-parametric case} Let us call by a {\it point of the configuration space} a set of smooth functions 
$\varphi(s)=(\varphi^i(s))$, $1\le i\le M$, $s=(s^1,\ldots,s^n)$. Let $\CC:x=x(s)$ be an $n$-dimensional
parameterized surface in the spacetime $\R^{n+1}$. Let us call by a {\it tangent element} to a point $\varphi(s)$ of the
configuration space, corresponding to the surface $\CC$, or by {\it Cauchy data} for Euler--Lagrange equations (\ref{Euler_Lagrange}), a set of functions $\varphi_x(s)=\varphi^i_{x^j}(s)$, $1\le i\le M$, $0\le j\le n$,
such that
\begin{equation}
\label{tan}
\sum_{j=0}^n \varphi^i_{x^j}(s)x^j_{s^k}(s)=\varphi^i_{s^k}(s),\ \ i=1,\ldots,M,\ \ k=1,\ldots,n.
\end{equation}

Let us call by the {\it integral element} corresponding to a surface $\CC$, to a point $\varphi(s)$, and to a tangent element $\varphi_x(s)$,
the set of functions $\pi_i(s)$, $H_j(s)$, $1\le i\le M$, $0\le j\le n$, given by (\ref{pi,H}). 

Since for a point $\varphi(s)$ and for a surface $\CC$ the tangent element $\varphi_x(s)$ subject to (\ref{tan}) at any $s\in\CC$ depends on $M(n+1)-Mn=M$ free parameters,
the transform $\varphi_x(s)\to\pi(s)=(\pi_i(s))$, $1\le i\le M$, is, in general, one-to-one. We shall call the space of smooth functions $(\varphi^i(s),\pi_i(s))$ by the {\it phase space}, and $\pi_i(s)$ by the {\it conjugate variable} to $\varphi^i(s)$.

In particular, we can, in general, express $H_j$ as functions 
\begin{equation}
\label{Ham_density}
H_j(s)=H_j(x(s),x_{s^k}(s),\varphi^i(s),\varphi^i_{s^k}(s),\pi_i(s)).
\end{equation}
We shall call $H_j(s)$ by the {\it $j$-th Hamiltonian density}.

Any solution $\varphi(x)$ of Euler--Lagrange equations (\ref{Euler_Lagrange}) yields, for any surface $\CC: x=x(s)$, a point $\varphi(s)=\varphi(x(s))$ of the configuration space
and a tangent element $\varphi_x(s)=\varphi_x(x(s))$, hence a point $(\varphi^i(\cdot),\pi_i(\cdot))$ of the phase space. Let 
$x(s,t)$ be a smooth function of $s=(s^1,\ldots,s^n)$ and of a parameter $t$. It yields a one-parametric family $(\varphi^i(\cdot,t),\pi_i(\cdot,t))$ of points
of the phase space. 

In the variables $t,\varphi^i(s),\pi_i(s)$ Euler--Lagrange equations (\ref{Euler_Lagrange}) become the following
{\it generalized canonical Hamilton equations}, for any function $x(s,t)$,
\begin{equation}
\label{Hamilton_equations}
\left\{
\begin{array}{l}
\frac{\partial\varphi^i}{\partial t}(s,t)=\frac{\delta H}{\delta\pi_i(s)}(t,\varphi^i(\cdot,t),\pi_i(\cdot,t)),\\ 
\frac{\partial\pi_i}{\partial t}(s,t)=-\frac{\delta H}{\delta\varphi^i(s)}(t,\varphi^i(\cdot,t),\pi_i(\cdot,t)), 
\end{array}
\right.
\end{equation}
where 
\begin{equation}
\label{Hamiltonian}
H=H(t,\varphi^i(\cdot),\pi_i(\cdot))=\int\sum_{j=0}^n x^j_t(s,t)H_j(s,x(\cdot,t),\varphi^i(\cdot),\pi_i(\cdot))ds,
\end{equation}
and $\delta/\delta\pi_i(s)$, $\delta/\delta\varphi^i(s)$ are functional derivatives, see e.~g. [\ref{St07}].
For a proof of (\ref{Hamilton_equations}), see \S\ref{Hamilton_formalism_parametric} below.

\subsubsection{Parametric case} 
\label{Hamilton_formalism_parametric} Let us call by a {\it point of the extended configuration space} a 
parameterized $n$-dimensional surface $C$ in $\R^N$, i.~e. a set of smooth functions $C=(y(s))=(y^l(s))$, $1\le l\le N$, $s=(s^1,\ldots,s^n)$, with the nondegenerate Jacobi matrix 
at any point $s\in C$ . Let us call  
by a {\it tangent element} to a point $C$ of the extended configuration space, or by {\it Cauchy data} for Euler--Lagrange equations (\ref{param_Euler_Lagrange}),
a smooth family of $(n+1)$-dimensional (oriented) planes in $\R^N$ tangent to the surface $C$ at each $s\in C$,
i.~e., for any nondegenerate $n\times(n+1)$-matrix of functions $u_s(s)=(u^j_{s^k}(s))$, $0\le j\le n$, $1\le k\le n$, a set of functions $y_u(s)=(y^l_{u^j}(s))$, $1\le l\le N$, $0\le j\le n$, such that
\begin{equation}
\label{yus}
\sum_{j=0}^n y^l_{u^j}(s)u^j_{s^k}(s)=y^l_{s^k}(s),\ \ 1\le l\le N,\ \ 1\le k\le n,
\end{equation}  
compatible with changes of functions 
\begin{equation}
\label{uw}
u_s(s)\mapsto w_s(s)=(w^j_{s_k}(s))=\left(\sum_{j'=0}^n w^j_{u^{j'}}(s)u^{j'}_{s^k}(s)\right)
\end{equation}
given by a nondegenerate $(n+1)\times(n+1)$-matrix $(w^j_{u^{j'}})_{0\le j,j'\le n}(s)$ with positive determinant, so that
\begin{equation}
\label{yuw}
y^l_{u^{j'}}(s)=\sum_{j=0}^n y^l_{w^j}(s)w^j_{u^{j'}}(s).
\end{equation}

Let us call by the {\it integral element} corresponding to a point $C: y^l=y^l(s)$, to a matrix $u_s(s)$, and to a tangent element $y_u(s)$,
the set of functions $v_l(s)$, $1\le l\le N$, given by (\ref{v_l}). It is easy to check that the integral element $v(s)=(v_l(s))$ is compatible with changes (\ref{uw}, \ref{yuw}), i.~e. does not depend on $u$ 
but depends only on the tangent element. 

We shall call the space of smooth functions $(y^l(s),v_l(s))$ by the {\it extended phase space}, and $v_l(s)$ by the {\it conjugate variable} to $y^l(s)$ .

The tangent element at any $s\in C$ depends on $N-n-1$ free parameters. Hence the variables $v_l(s)$ satisfy $n+1$ equations. $n$ of these equations are easy to find:
\begin{equation}
\label{HJ_vector_fields}
\HH^k(s)\equiv\HH^k(y^l(s),y^l_s(s),v_l(s))\equiv\sum_{l=1}^N v_l(s) y^l_{s^k}(s)=0,\ \  k=1,\ldots,n.
\end{equation}
These equations express the fact that integral (\ref{variation_of_parametric_action}) is independent on the choice of parameters $s$ on the boundary surface $\CC$.

The remaining $(n+1)$-th equation depends on the Lagrangian $\LL$. Denote this equation by
\begin{equation}
\label{HJ_constraint}
\HH^0(s)\equiv\HH^0(y^l(s),y^l_{s^k}(s),v_l(s))=0.
\end{equation}
We shall call equations (\ref{HJ_vector_fields}, \ref{HJ_constraint}) by the {\it generalized Hamilton--Jacobi constraints}.  

For non-parametric case, equations (\ref{HJ_vector_fields}, \ref{HJ_constraint}) are equivalent to equations (\ref{Ham_density}) with 
\begin{equation}
\label{nonparam}
(y^l)=(x^j,\varphi^i),\ \ (v_l)=(-H_j,\pi_i).
\end{equation}
In these notations, equations (\ref{HJ_vector_fields}) are
\begin{equation}
\label{nonparam_HJ_k}
\HH^k(s)\equiv\sum_{i=1}^M\pi_i\varphi^i_{s^k}-\sum_{j=0}^n H_j x^j_{s^k}=0,\ \ k=1,\ldots,n.
\end{equation}

{\it Examples.} 1) The scalar self-interacting field, see (\ref{scalar_field}) above. In this case, in notations (\ref{nonparam}), we have [\ref{St07}]
\begin{equation}
\label{HJ_scalar_field}
\begin{aligned}
&\HH^0(s)=-\sum_\mu D_\mu H_\mu(s)+\sum_\mu D_\mu D^\mu V(\varphi(s))\\
&+\frac 12\pi(s)^2-\frac 12 \sum\limits_{\mu<\nu}\sum\limits_{k,k'}D^{\mu\nu}_k D_{\mu\nu,k'}\varphi_{s^k}\varphi_{s^{k'}},
\end{aligned}
\end{equation}
where we have introduced Greek indices $\mu, \nu$ instead of $j$, raising and lowering Greek indices goes using the
Lorentz metric
\begin{equation}
dx^2=(dx^0)^2-\sum_{j\ne 0}(dx^j)^2,
\end{equation}
and
\begin{equation}
\label{D^mu^nu}
\begin{aligned}
 &D^\mu=(-1)^\mu \frac{\partial(x^0,\ldots,\widehat{x^\mu},\ldots,x^n)}{\partial(s^1,\ldots,s^n)},\\
D^{\mu\nu}_k&=(-1)^{k+\mu+\nu}\frac{\partial(x^0,\ldots,\widehat{x^\mu},\ldots,\widehat{x^\nu},\ldots,x^n)}
{\partial(s^1,\ldots,\widehat{s^k},\ldots,s^n)}.
\end{aligned}
\end{equation}

2) The Dirichlet principle, see (\ref{Dirichlet}) above. In this case we have 
\begin{equation}
\label{HJ_Dirichlet}
\HH^0(s)=-x^1_s H_0(s)+x^0_s H_1(s)+\frac12\pi(s)^2-\frac12\varphi_s^2.
\end{equation}

3) The Plateau problem, see (\ref{Plateau}) above. In this case we have [\ref{St07}]
\begin{equation}
\label{HJ_Plateau}
\HH^0(s)=\sum_{l=1}^N v_l(s)^2-\sum_{1\le l_1<\ldots<l_n\le N}\left(\frac{\partial(y^{l_1},\ldots,y^{l_n})}{\partial(s^1,\ldots,s^n)}\right)^2.
\end{equation}
 
Note that function $\HH^0(s)$ is formally defined not uniquely. For instance, in the latter Example one can also put
\begin{equation}
\label{HJ2_Plateau}
\HH^0(s)=\sqrt{\sum_{l=1}^N v_l(s)^2}-\sqrt{\sum_{1\le l_1<\ldots<l_n\le N}\left(\frac{\partial(y^{l_1},\ldots,y^{l_n})}{\partial(s^1,\ldots,s^n)}\right)^2}+\sum_{k=1}^n\HH^k(s).
\end{equation}

Any solution $y(u)$ of Euler--Lagrange equations (\ref{param_Euler_Lagrange}) yields, for any function $u(s)$, a 
point $C=(y(s))=y(u(s))$ of the extended configuration space and a 
tangent element $y_u(s)=y_u(u(s))$ compatible with changes (\ref{uw}, \ref{yuw}), hence a point $(y(\cdot),v(\cdot))$ of the extended phase space.
Let $u(s,t)$ be a smooth function of $s=(s^1,\ldots,s^n)$ and of a parameter $t$. It yields a one-parametric family $(y^l(\cdot,t),v_l(\cdot,t))$ of points
of the extended phase space. 

\begin{theorem}
\label{theorem_param_Hamilton}
In the variables $t,y^l(s),v_l(s)$ Euler--Lagrange equations \emph{(\ref{param_Euler_Lagrange})} become the following parametric generalized
canonical Hamilton equations, for any function $u(s,t)$,
\begin{equation}
\label{param_Hamilton}
\left\{\begin{array}{l}
\frac{\partial y^l}{\partial t}(s,t)=\frac{\delta H}{\delta v_l(s)}(t,y^l(\cdot,t),v_l(\cdot,t)),\\
\frac{\partial v_l}{\partial t}(s,t)=-\frac{\delta H}{\delta y^l(s)}(t,y^l(\cdot,t),v_l(\cdot,t)),\\
H=H(t,y^l(\cdot),v_l(\cdot))=\int\sum_{j=0}^n\lambda_j(s,t)\HH^j(s,y^l(\cdot),v_l(\cdot))ds,\\
\HH^j(s,y^l(\cdot,t),v_l(\cdot,t))=0,
\end{array}\right.
\end{equation}
for certain functions $\lambda_j(s,t)$, $j=0,\ldots,n$, depending on $u(s,t)$, uniquely determined from system \emph{(\ref{param_Hamilton})}.
\end{theorem}

{\it Proof.} Consider the variational principle for the integral
\begin{equation}
\label{Hilbert}
\int\!\int\sum_{l=1}^N v_l(s,t)y^l_t(s,t)ds\,dt
\end{equation}
for arbitrary functions $y^l(s,t),v_l(s,t)$ subject to constraints 
\begin{equation}
\label{constraint_Hilbert}
\HH^j(s,y^l(\cdot,t),v_l(\cdot,t))=0,\ \  j=0,\ldots,n. 
\end{equation}

Let us fix $s$, $t$, $y^l$, and $y^l_{s^k}$. 

\begin{lemma} 
\label{duality}
For given $y^l_t$, extremum of the sum $\sum_{l=1}^N v_l y^l_t$
for $v_l$ subject to \emph{(\ref{constraint_Hilbert})} is achieved at 
\begin{equation}
\label{v_l_from_y^l_t} 
v_l=\LL_{y^l_t}(y^l,y^l_{s^k},y^l_t),\ \  \sum_{l=1}^N v_l y^l_t=\LL(y^l,y^l_{s^k},y^l_t).
\end{equation}
\end{lemma}

{\it Proof.} By the Lagrange multipliers method, this extremum is achieved at $v_l$ such that
\begin{equation}
\label{y^l_t_from_v_l}
y^l_t=\sum_{j=0}^n \lambda_j\HH^j_{v^l}=\sum_{k=1}^n\lambda_k y^l_{s^k}+\lambda_0\HH^0_{v^l}
\end{equation}
for some numbers $\lambda_0,\ldots,\lambda_n$.

Consider the quotient vector space $U$ of the vector space $\R^N$ of vectors $y^l_t$ by the subspace spanned by the vectors $y^l_{s^k}$, $k=1,\ldots,n$.
Function $\LL=\LL(y^l,y^l_{s^k},y^l_t)$, as a function of $y^l_t$, is a homogeneous function of degree 1 on $U$. The dual vector space $U'$ consists of vectors $v_l$ such that 
$\HH^k(v_l)=\sum_{l=1}^N v_l y^l_{s^k}=0$, $1\le k\le n$. Since $v_l$ in parameters $u$ and in parameters $(s,t)$
are the same, formulas (\ref{v_l}) and (\ref{v_l_from_y^l_t}) for $v_l$ are equivalent. We have a map $U\to U'$ given by (\ref{v_l_from_y^l_t}). The image of this map coincides with 
the set of vectors $v_l$ such that $\HH^0(v_l)=0$. Now equivalence of (\ref{v_l_from_y^l_t}) and (\ref{y^l_t_from_v_l}) is a well known fact called projective duality on the space $U$,
or Legendre transform on $U$, or duality of finite dimensional normed spaces $U$ and $U'$, see e.~g. [\ref{St07}]. Q.~E.~D.
\medskip

{\it End of proof of Theorem \emph{\ref{theorem_param_Hamilton}}.}
By Lemma \ref{duality}, variational principle (\ref{Hilbert}, \ref{constraint_Hilbert}) is equivalent to variational principle (\ref{parametric_action}).

On the other hand, by the Lagrange multipliers method, variational principle (\ref{Hilbert}, \ref{constraint_Hilbert}) is equivalent to 
the variational principle for the integral  
\begin{equation}
\int\!\int\left(\sum_{l=1}^N v_l(s,t)y^l_t(s,t)-\sum_{j=0}^n\lambda_j(s,t)\HH^j(s,y^l(\cdot,t),v_l(\cdot,t))\right)ds\,dt
\end{equation}
for arbitrary functions $y^l(s,t),v_l(s,t)$. Writing the Euler--Lagrange equations for this variational principle, we obtain system (\ref{param_Hamilton}). 

Theorem \ref{theorem_param_Hamilton} is proved.
\medskip

Note that in derivation of equations (\ref{param_Hamilton}) we did not use the concrete form (\ref{HJ_vector_fields}, \ref{HJ_constraint}) of equations $\HH^j(s)$. Hence we
can apply this argument to equations (\ref{Ham_density}). This way we obtain non-parametric generalized canonical Hamilton equations (\ref{Hamilton_equations}, \ref{Hamiltonian}).

\subsection{Classical observables}

\subsubsection{Non-parametric case}
\label{observ_nonparametric}
Let us call by a {\it classical observable} a functional
\begin{equation}
\Phi(x^j(\cdot),\varphi^i(\cdot),\varphi^i_{x^j}(\cdot))=\Phi(x^j(\cdot),\varphi^i(\cdot),\pi_i(\cdot)) 
\end{equation}
of functions $x^j(s)$, $\varphi^i(s)$, and $\varphi^i_{x^j}(s)$ subject to (\ref{tan}) or $\pi_i(s)$,
which is a first integral of Euler--Lagrange equations (\ref{Euler_Lagrange}) or generalized canonical Hamilton equations (\ref{Hamilton_equations}), i.~e. depends not on 
$x(s), \varphi(s), \varphi_x(s)$ or $\pi(s)$ but only on a solution 
$\varphi(x)$ of the Euler--Lagrange equations. 
In the variables $x^j(s),\varphi^i(s),\pi_i(s)$ this can be written in the form
\begin{equation}
\label{equation_for_nonparametric_observables}
\frac{\delta\Phi}{\delta x^j(s)}+\{H_j(s),\Phi\}=0,
\end{equation}
where
\begin{equation}
\label{Poisson_bracket}
\{\Phi_1,\Phi_2\}=\int\sum_{i=1}^M\left(\frac{\delta\Phi_1}{\delta\pi_i(s)}\frac{\delta\Phi_2}{\delta\varphi^i(s)}-\frac{\delta\Phi_1}{\delta\varphi^i(s)}\frac{\delta\Phi_2}{\delta\pi_i(s)}\right)\,ds
\end{equation}
is the Poisson bracket of two functionals 
$$
\Phi_1(x^j(\cdot),\varphi^i(\cdot),\pi_i(\cdot))\text{ and }\Phi_2(x^j(\cdot),\varphi^i(\cdot),\pi_i(\cdot)).
$$

Functions $H_j(s)$ satisfy the Frobenius integrability (zero curvature) condition,
\begin{equation}
\label{Frob}
\frac{\delta H_j(s)}{\delta x^{j'}(s')}-\frac{\delta H_{j'}(s')}{\delta x^j(s)}-\{H_j(s),H_{j'}(s')\}=0.
\end{equation}
For a proof, see \S\ref{observ_parametric} below.
This condition means that solution of system (\ref{equation_for_nonparametric_observables}) is well defined.

Observables form a commutative Poisson algebra, to be denoted $A_0$, with respect to product of functionals and Poisson bracket (\ref{Poisson_bracket}).

\subsubsection{Parametric case}
\label{observ_parametric}

Let us call by a {\it classical \emph(parametric\emph) observable} a functional
$$
\Phi(y(\cdot),y_u(\cdot),u_s(\cdot))=\Phi(y(\cdot),v(\cdot))
$$ 
of functions $y(s), y_u(s), u_s(s)$ subject to (\ref{yus}) or $v(s)$, 
which is a density in $u$ of degree 0, cf. (\ref{density_LL_BB}), i.~e. depends not on $u$ but only on the tangent element, and which is a first integral of Euler--Lagrange equations 
(\ref{param_Euler_Lagrange}) or parametric generalized canonical Hamilton equations (\ref{param_Hamilton}), i.~e. depends not on 
$y(s), y_u(s), u_s(s)$ or $v(s)$ but only on a solution $y(u)$ of the Euler--Lagrange equations. 
In the variables $y^l(s),v_l(s)$ this can be written in the form, for any function $u(s,t)$,
\begin{equation}
\label{param_equation_for_observables}
\Phi'=\{H(t),\Phi\}=0,
\end{equation}
where
\begin{equation}
\label{param_Poisson_bracket}
\{\Phi_1,\Phi_2\}=\int\sum_{l=1}^N\left(\frac{\delta\Phi_1}{\delta v_l(s)}\frac{\delta\Phi_2}{\delta y^l(s)}-\frac{\delta\Phi_1}{\delta y^l(s)}\frac{\delta\Phi_2}{\delta v_l(s)}\right)\,ds
\end{equation}
is the Poisson bracket of two functionals $\Phi_1(y(\cdot),v(\cdot))$ and $\Phi_2(y(\cdot),v(\cdot))$, and $H(t)$ is a continual linear combination of $\HH^j(s)$, see (\ref{param_Hamilton}). 

Note that parametric generalized canonical Hamilton equations (\ref{param_Hamilton}) preserve constraints $\HH^j(s)=0$. This means that
\begin{equation}
\label{Poisson_H_constraints}
\{H(t),\HH^j(s)\}=\int\sum_{j'=0}^n \Psi^j_{j'}(s,s')\HH^{j'}(s')\,ds'
\end{equation}
for some functionals $\Psi^j_{j'}(s,s')=\Psi^j_{j'}(s,s',y^l(\cdot),v_l(\cdot))$.

When function $u(s,t)$
varies, $H(t)$, in general, spans all continual linear combinations of $\HH^j(s)$. Therefore we can rewrite (\ref{param_equation_for_observables}) as
\begin{equation}
\label{param_bracket_observables}
\{\HH^j(s),\Phi\}=0,
\end{equation}
and we can rewrite (\ref{Poisson_H_constraints}) as
\begin{equation}
\label{Poisson_constraints}
\{\HH^j(s),\HH^{j'}(s')\}=\int\sum_{j''=0}^n\Psi^{j,j'}_{j''}(s,s',s'')\HH^{j''}(s'')\,ds''
\end{equation}
for some functionals $\Psi^{j,j'}_{j''}(s,s',s'')=\Psi^{j,j'}_{j''}(s,s',s'',y^l(\cdot),v_l(\cdot))$.

Equalities (\ref{param_bracket_observables}) and (\ref{Poisson_constraints}) do not depend on concrete form of equations $\HH^j(s)$. Hence we can apply them for equations (\ref{Ham_density}). 
Then (\ref{param_bracket_observables}) becomes equation (\ref{equation_for_nonparametric_observables}) for non-parametric classical observables, and (\ref{Poisson_constraints}) becomes  Frobenius 
integrability condition (\ref{Frob}).

Observables form a commutative Poisson algebra with respect to product of functionals and Poisson bracket (\ref{param_Poisson_bracket}). For 
nonparametric case, this algebra coincides with the algebra $A_0$ of nonparametric classical observables constructed in \S\ref{observ_nonparametric} above, and Poisson bracket (\ref{param_Poisson_bracket}) coincides 
with non-parametric Poisson bracket (\ref{Poisson_bracket}).
Hence we denote the Poisson algebra of parametric classical observables also by $A_0$, 
without abuse of notation. This Poisson algebra $A_0$ can be described 
as follows. Consider the Poisson algebra $\Delta_0$ of functionals $\Phi(y(\cdot),v(\cdot))$ on the extended phase space with Poisson bracket (\ref{param_Poisson_bracket}). Then one has
\begin{equation}
\label{A_0}
A_0\simeq E_0/F_0,
\end{equation}
where 
\begin{equation}
\label{F_0}
F_0=\langle\Delta_0\HH^j(s)\rangle
\end{equation}
is the ideal in $\Delta_0$ generated by $\HH^j(s)$, and
\begin{equation}
\label{E_0}
E_0=\left(\Phi\in \Delta_0:\{\Pi,\Phi\}\in F_0\text{ for any }\Pi\in F_0\right).
\end{equation}
Inclusion $F_0\subset E_0$ follows from (\ref{Poisson_constraints}). It means that $F_0$ is a Poisson ideal in $\Delta_0$, i.~e. an ideal closed with respect to Poisson bracket.
Construction (\ref{A_0}, \ref{E_0}) yields a Poisson algebra for any Poisson algebra $\Delta_0$ and a Poisson ideal $F_0$ in $\Delta_0$.
This construction is a generalization and simplification of the construction of Hamiltonian reduction, which is obtained if the Poisson ideal $F_0$ is generated by a Lie subalgebra of $\Delta_0$ with respect to 
Poisson bracket. For instance, this is the case if coefficients $\Psi^{j,j'}_{j''}(s,s',s'')\in\Delta_0$ in (\ref{Poisson_constraints}) are constant functionals. 
In particular, this holds for nonparametric Hamilton--Jacobi constraints (\ref{Ham_density}) instead of $\HH^j(s)$. In this case the Lie algebra spanned by the constraints is commutative.

Another example is the Lie algebra spanned by the constraints $\HH^k(s)$ (\ref{HJ_vector_fields}), $1\le k\le n$. It is the Lie algebra $\Vect_n$ of vector fields in the
variables $s$. The algebra $A_0$ (\ref{A_0}--\ref{E_0}) is obtained from $\Delta_0$ by the following two step construction.

{\it Step} 1: imposing the constraint $\HH^0(s)$. The result of this step is the Poisson algebra
$\Gamma_0=E_0/F_0$, where $F_0$ is the Poisson ideal in $\Delta_0$ generated by $\HH^0(s)$, and $E_0$ is given by (\ref{E_0}). 

{\it Step} 2: Hamiltonian reduction with respect to the Lie algebra $\Vect_n$. This step is especially simple if $\HH^0(s)$ is covariant with respect to the Lie algebra $\Vect_n$ of vector fields, as it is,
for example, for scalar self-interacting field (\ref{HJ_scalar_field}) and for Plateau problem (\ref{HJ_Plateau}). In this case we have $\HH^k(s)\in\Gamma_0$, and the result of 
reduction is the Poisson algebra 
\begin{equation}
\label{A_0_from_Gamma_0}
A_0=(\Gamma_0/\langle\Gamma_0\Vect_n\rangle)^{\Vect_n}
\end{equation}
of $\Vect_n$-invariant elements in the quotient space of the algebra $\Gamma_0$  by the Poisson ideal $\langle\Gamma_0\Vect_n\rangle$ generated by $\Vect_n\subset\Gamma_0$.

\subsection{Generalized Noether theorem}
\label{subsect_Noether}
Here we find the role of symmetries in the theory.

\subsubsection{Parametric case}

Let $\g$ be the Lie algebra of the symmetry Lie group $G$ of parametric action (\ref{parametric_action}). Let $\alpha=\delta P(\varepsilon)/\delta\varepsilon|_{\varepsilon=0}\in\g$, 
where $P(\varepsilon)$ is a curve in $G$ with $P(0)=1$. Put $u=(s,t)$, $y(u,\varepsilon)=P(\varepsilon)y(u)$. Then variation (\ref{variation_of_parametric_action}) of parametric action vanishes.
Differentiating equality (\ref{variation_of_parametric_action}) for $\LL=\LL(y,y_s,y_t)$ and $\BB=\BB(y,y_s,y_t)$ with respect to $t$, we obtain
\begin{equation}
\begin{aligned}
&\sum_{l=1}^N\left[\LL_{y^l}\delta y^l(y)+\sum_{k=1}^n \LL_{y^l_{s^k}}\delta y^l_{s^k}(y,y_s)+\LL_{y^l_t}\delta y^l_t(y,y_t)\right]\\
&+\sum_{l=1}^N\left[(\delta\BB)_{y^l}y^l_t+\sum_{k=1}^n(\delta\BB)_{y^l_{s^k}}y^l_{s^k t}+(\delta\BB)_{y^l_t}y^l_{tt}\right]=0
\end{aligned}
\end{equation}
for any $y^l,y^l_{s^k},y^l_t,y^l_{s^k t},y^l_{tt}$. This impies that:

(i) $(\delta\BB)_{y^l_{s^k}}=(\delta\BB)_{y^l_t}=0$, i.~e. $\delta\BB(y,y_s,y_t)=\delta\BB(y)$ does not depend on $y_s, y_t$;

(ii) for
\begin{equation}
\label{def_I}
I(y(\cdot),v(\cdot))\delta\varepsilon=\int\left[\sum_{l=1}^N v_l(s)\delta y^l(y(s))+\delta\BB(y(s))\right]\,ds,
\end{equation}
we have 
\begin{equation}
\frac{dI}{dt}\delta\varepsilon=\int\sum_{l=1}^N\left(-\LL_{y^l}+\sum_{k=1}^n\frac\partial{\partial s^k}\LL_{y^l_{s^k}}+\frac\partial{\partial t}\LL_{y^l_t}\right)\delta y^l(y(s))ds=0,
\end{equation}
i.~e. $I(y(\cdot),v(\cdot))=I_\alpha(y(\cdot),v(\cdot))$ is a first integral, 
\begin{equation}
\label{Noether_1integral}
\{\HH^j(s),I\}=0.
\end{equation}

\begin{theorem} 
\label{theorem_Noether}
\emph{(i)} For any functional $\Phi(y(\cdot),v(\cdot))$, we have 
\begin{equation}
\label{Noether_var}
\delta\Phi/\delta\varepsilon=\{I,\Phi\}.
\end{equation}

\emph{(ii)} For $\alpha,\beta\in\g$, we have 
\begin{equation}
\label{homom}
I_{\alpha+\beta}=I_{\alpha}+I_{\beta},\ \ I_{[\alpha,\beta]}=\{I_\alpha,I_\beta\}.
\end{equation}
\end{theorem}

{\it Proof.} (i) It suffices to check (\ref{Noether_var}) for $\Phi=y^l(s)$ and for 
\begin{equation}
\Phi=y^l_t(s)=\sum_{j=0}^n\lambda_j\HH^j_{v_l}(s), 
\end{equation}
see (\ref{y^l_t_from_v_l}).
For $\Phi=y^l(s)$, (\ref{Noether_var}) is obvious.
For $\Phi=y^l_t(s)$, we have 
\begin{equation}
\begin{aligned}
&\delta y^l_t(s)=(\delta y^l)_t(s)=\sum_{l'=1}^N\frac{\partial\delta y^l}{\partial y^{l'}}(y(s))y^{l'}_t(s)\\
&=\sum_{l'=1}^N\frac{\partial\delta y^l}{\partial y^{l'}}(y(s))\sum_{j=0}^n\lambda_j\HH^j_{v_{l'}}(s)
=\left\{I,\sum_{j=0}^n\lambda_j\HH^j_{v_l}(s)\right\}\delta\varepsilon.
\end{aligned}
\end{equation}
The latter equality is obtained by differentiating equality (\ref{Noether_1integral}) with respect to $v_l(s')$. Q.~E.~D. 

(ii) directly follows from (\ref{def_I}). Q.~E.~D.

\begin{theorem}
\label{theorem_canon}
For any $g\in G$, the transform 
\begin{equation}
\label{canon}
(y(\cdot),v(\cdot))\mapsto (\tilde y(\cdot),\tilde v(\cdot))=g(y(\cdot),v(\cdot)),
\end{equation}
arising from the transform $g$ of tangent elements in the extended configuration space,
preserves Poisson bracket \emph{(\ref{param_Poisson_bracket})}.
\end{theorem}

{\it Proof.} For a connected Lie group $G$, the statement follows from Theorem~\ref{theorem_Noether}. For a general Lie group $G$ one can argue as follows.
 Formula (\ref{variation_of_parametric_action}) for variation of parametric action implies the equality of differential 1-forms on the extended phase space
\begin{equation}
\begin{aligned}
&\int\sum_{l=1}^N\tilde v_l(s)\delta\tilde y^l(s)\,ds+\delta\int\BB(\tilde y(s),\tilde v(s))\,ds\\
&=\int\sum_{l=1}^N v_l(s)\delta y^l(s)\,ds+\delta\int\BB(y(s),v(s))\,ds.
\end{aligned}
\end{equation}
Taking the exterior differential, we obtain
\begin{equation}
\int\sum_{l=1}^N\delta\tilde v_l(s)\wedge\delta\tilde y^l(s)\,ds=\int\sum_{l=1}^N\delta v_l(s)\wedge\delta y^l(s)\,ds,
\end{equation}
i.~e. $(y(\cdot),v(\cdot))\mapsto(\tilde y(\cdot),\tilde v(\cdot))$ is a canonical transform. Hence this transform preserves Poisson bracket (\ref{param_Poisson_bracket}). Q.~E.~D.
\medskip

Summarizing, we obtain that the Poisson algebra $A_0$ of classical parametric observables is a {\it $G$-equivariant Poisson algebra}, i.~e. it is a Poisson algebra equipped with
a Lie group $G$ of automorphisms and with a homomorphism of Lie algebras $\g\to A_0$, $\alpha\mapsto I_\alpha$, such that the infinitesimal action of $\alpha\in\g$ 
on $A_0$ arising from the $G$-action is given by the Poisson bracket with $I_\alpha$.   

\subsubsection{Non-parametric case} Transferring the latter results to non-par\-am\-etric case, we obtain the following. Let $\g$ be the Lie algebra of the symmetry Lie group $G$ of action (\ref{action}).
Then we have a Lie algebra homomorphism $\g\to A_0$, $\alpha\mapsto I=I_\alpha$, where
\begin{equation}
\label{nonparam_1integral}
\begin{aligned}
&I(x(\cdot),\varphi(\cdot),\pi(\cdot))\delta\varepsilon=\int\left[\sum_{i=1}^M \pi_i(s)\delta\varphi^i(x(s),\varphi(s))\right.\\
&\left.-\sum_{j=0}^n H_j(s,x(\cdot),\varphi(\cdot),\pi(\cdot))\delta x^j(x(s),\varphi(s))+\delta B(x(s),\varphi(s))\right]\,ds,
\end{aligned}
\end{equation}
such that for any functional $\Phi(x(\cdot),\varphi(\cdot),\pi(\cdot))$ we have equality (\ref{Noether_var}). Also the $G$-action preserves Poisson bracket (\ref{Poisson_bracket}). In other words, $A_0$ is a $G$-equivariant Poisson algebra.

\section{Quantization of classical field theories}

\subsection{Non-parametric case}
\label{nonparam_quant}

We are going to construct a quantization of the $G$-equivariant commutative Poisson algebra $A_0$, i.~e. a $G$-equi\-var\-iant complex associative algebra $A_h$ of quantum observables
$\Phi^{(h)}$ with associative product $\Phi_1^{(h)}*\Phi_2^{(h)}=\Phi_1^{(h)}*_h\Phi_2^{(h)}$, smoothly depending on the real parameter $h$ (the Planck constant),
such that classical observables $\Phi^{(0)}$ coincide with complex valued solutions of equation (\ref{equation_for_nonparametric_observables}), 
and one has
\begin{equation}
\label{quantization}
\begin{aligned}
\Phi_1^{(0)}*_0\Phi_2^{(0)}&=\Phi_1^{(0)}\Phi_2^{(0)};\\ 
\lim_{h\to 0}\frac ih[\Phi_1^{(h)},&\Phi_2^{(h)}]=\{\Phi_1^{(0)},\Phi_2^{(0)}\},\\ 
\end{aligned}
\end{equation}
where 
\begin{equation}
\label{commutator}
[\Phi_1,\Phi_2]=\Phi_1*\Phi_2-\Phi_2*\Phi_1
\end{equation}
is the commutator, and the second limit equality in (\ref{quantization}) holds for arbitrary smooth maps $h\mapsto\Phi_1^{(h)}$, $h\mapsto\Phi_2^{(h)}$.  
$G$-equivariance means that $A_h$ is an algebra 
equip\-ped with a Lie group $G$ of automorphisms and with a homomorphism of Lie algebras $\g\to A_h$, $\alpha\mapsto i\widehat I_\alpha/h=iI_\alpha^{(h)}/h$, 
\begin{equation}
\label{properties_I_alpha}
\widehat I_{[\alpha,\beta]}=\frac ih[\widehat I_\alpha,\widehat I_\beta],
\end{equation}
such that the infinitesimal action of $\alpha\in\g$ on $A_h$ arising from the $G$-action is given by the commutator with $i\widehat I_\alpha/h$.

The usual way to construct such a quantization is the following. First, one constructs a quantization of the Poisson algebra of functionals 
$\Phi(\varphi(\cdot),\pi(\cdot))$ on the phase space with Poisson bracket (\ref{Poisson_bracket}). Denote this quantization by $\Delta^0_h$, with the product $*_h$. 
One usually realizes the algebra $\Delta^0_h$ as an algebra of operators on a vector space $W_0$. Next, consider the generalized Schr\"odinger equation
\begin{equation}
\label{gen_Schrod}
ih\delta\Psi(x(\cdot))/\delta x^j(s)=\widehat H_j(s,x(\cdot))\Psi(x(\cdot))
\end{equation} 
for $\Psi(x(\cdot))\in W_0$, where $\widehat H_j(s,x(\cdot))=H_j(s,x(\cdot))^{(h)}\in \Delta^0_h$ is an appropriate quantization of the Hamiltonian density $H_j(s$, $x(\cdot)$, $\varphi(\cdot)$, $\pi(\cdot))$. 
The algebra $A_h$ is defined as the algebra of operators $\Phi^{(h)}(x(\cdot))\in \Delta^0_h$
preserving the space of solutions $\Psi(x(\cdot))$ of the generalized Schr\"odinger equation. This means that $\Phi^{(h)}(x(\cdot))$ satisfies the generalized Heisenberg equation
\begin{equation}
\label{gen_Heisenberg}
ih\delta\Phi^{(h)}(x(\cdot))/\delta x^j(s)=[\widehat H_j(s,x(\cdot)),\Phi^{(h)}(x(\cdot))].
\end{equation}
The space of solutions $\Phi^{(h)}(x(\cdot))$ of the generalized Heisenberg equation is an associative algebra with respect to the product $*_h$, which is the required quantization of the Poisson algebra of classical observables 
$\Phi(x(\cdot),\varphi(\cdot),\pi(\cdot))=\Phi^{(0)}(x(\cdot))$ satisfyng equation (\ref{equation_for_nonparametric_observables}).

However, it turns out that this construction cannot be realized for general classical field theories, see e.~g. [\ref{St07}]. The problem is that the algebra $\Delta^0_h$ 
with reasonable properties of solutions of generalized Heisenberg equation (\ref{gen_Heisenberg}) does not exist.

In the present paper we propose a somewhat different construction, based on rewriting the Poisson algebra of classical observables in parametric form (\ref{A_0}--\ref{E_0}). 

\subsection{Parametric case} 
\label{param_quant}
First, consider a quantization $\Delta_h$ of the Poisson algebra $\Delta_0$ of functionals on the extended phase space. 
Let $\widehat\HH^j(s)=\HH^j(s)^{(h)}\in \Delta_h$ be an (appropriate) quantization of the generalized Ham\-ilt\-on--Jacobi constraint $\HH^j(s)\in \Delta_0$. 
We shall call $\widehat\HH^j(s)$ by the {\it \emph(parametric\emph) generalized Schr\"od\-ing\-er constraint}. 

One usually realizes the algebra $\Delta_h$ as an algebra of operators on a vector space $W$. Then we define $A_h$ as the algebra of operators from $\Delta_h$  
on the space of solutions $\Psi\in W$ of the generalized Schr\"odinger equation $\widehat\HH^j(s)(\Psi)=0$. For scalar field with self-action (\ref{nonparam_HJ_k}, \ref{HJ_scalar_field}), the 
generalized Schr\"odinger equation is a mathematical version of the Tomonaga--Schwinger equation known in QFT.

However, it turns out that the notion of solution of the generalized Schr\"odinger equation with reasonable properties cannot be defined for general classical field theories, see e.~g. [\ref{St07}]. 
Hence we propose a construction of the algebra $A_h$ not using the notion of solution. Put
\begin{equation}
\label{A_h}
A_h=E_h/F_h,
\end{equation}
where
\begin{equation}
\label{F_h}
F_h=\langle\Delta_h*\widehat\HH^j(s)\rangle
\end{equation}
is the left ideal in $\Delta_h$ generated by $\widehat\HH^j(s)$, and
\begin{equation}
\label{E_h}
E_h=\left(\Phi\in \Delta_h:\Pi*\Phi\in F_h\text{ for any }\Pi\in F_h\right).
\end{equation}
It is easy to see that $E_h$ is an associative algebra, and $F_h$ is a two-sided ideal in $E_h$. Hence $A_h$ is an associative algebra.
It is the required quantization of the Poisson algebra $A_0$ (\ref{A_0}--\ref{E_0}). Construction (\ref{A_h}, \ref{E_h}) yields an associative algebra for any associative algebra $\Delta_h$ and a 
left ideal $F_h$ in $\Delta_h$. This construction is a generalization and simplification of the construction of quantum Hamiltonian reduction, which is obtained if the left ideal $F_h$ is generated by a Lie subalgebra of $\Delta_h$
with respect to commutator (\ref{commutator}). For instance, let us mention the Lie algebra $\Vect_n$ of vector fields spanned by $\widehat\HH^k(s)$, $1\le k\le n$,
see below. 

We shall give an explicit construction of the algebras $\Delta_h$ and $A_h$. Define $\Delta_h$ as the Weyl algebra, i.~e. put
$\Delta_h=\Delta_0$ as a vector space, with the Moyal product
\begin{equation}
\label{Moyal}
\begin{aligned}
&\Phi_1*_h\Phi_2(y(\cdot),v(\cdot))=\\
&\exp\frac{ih}2\int\sum_{l=1}^N\left(-\frac\delta{\delta v_l^{(1)}(s)}\frac\delta{\delta {y^l}^{(2)}(s)}+\frac\delta{\delta {y^l}^{(1)}(s)}\frac\delta{\delta v_l^{(2)}(s)}\right)\,ds\\
&\Phi_1(y^{(1)}(\cdot),v^{(1)}(\cdot))\Phi_2(y^{(2)}(\cdot),v^{(2)}(\cdot))|_{y^{(1)}(\cdot)=y^{(2)}(\cdot)=y(\cdot),v^{(1)}(\cdot)=v^{(2)}(\cdot)=v(\cdot)}.
\end{aligned}
\end{equation}  

Define the generalized Schr\"odinger constraint as 
\begin{equation}
\label{Schrod}
\widehat\HH^j(s)=\HH^j(s)\in \Delta_h.
\end{equation}
Define the algebra $A_h$ by (\ref{A_h}--\ref{E_h}). Then the algebra $A_h$ is obtained from $\Delta_h$ by the following two step construction.

{\it Step} 1: imposing the constraint $\widehat\HH^0(s)$. The result of this step is the algebra
$\Gamma_h=E_h/F_h$, where $F_h$ is the left ideal in $\Delta_h$ generated by $\widehat\HH^0(s)$, and $E_h$ is given by (\ref{E_h}).  
The algebra $\Gamma_h$
is a quantization of the Poisson algebra $\Gamma_0$ constructed in \S\ref{observ_parametric}.

{\it Step} 2: quantum Hamiltonian reduction with respect to the Lie algebra $\Vect_n$ spanned by $\widehat\HH^k(s)$, $1\le k\le n$. 
This step is especially simple if $\widehat\HH^0(s)$ is covariant with respect to the Lie algebra $\Vect_n$ of vector fields, as it is,
for example, for scalar self-interacting field (\ref{HJ_scalar_field}) and for Plateau problem (\ref{HJ_Plateau}). 
In this case we have $\widehat\HH^k(s)\in\Gamma_h$, and the result of 
reduction is the algebra 
\begin{equation}
\label{A_h_from_Gamma_h}
A_h=(\Gamma_h/\langle\Gamma_h*\Vect_n\rangle)^{\Vect_n}
\end{equation}
of $\Vect_n$-invariant elements in the quotient space of the algebra $\Gamma_h$  by the left ideal $\langle\Gamma_h*\Vect_n\rangle$ generated by $\Vect_n\subset\Gamma_h$.  

Let us examine $G$-equivariance of this quantization $A_h$.

For $\alpha\in\g$, define 
\begin{equation}
\widehat I_\alpha(y,v)=I_\alpha(y,v)\in \Delta_h,
\end{equation}
where $I_\alpha(y,v)=I(y,v)$ is given by (\ref{def_I}). Then we have equalities (\ref{properties_I_alpha}).

Moreover, for Examples (\ref{HJ_scalar_field}) and (\ref{HJ_Plateau}) above, we have
\begin{equation}
\label{I_alpha_HH}
[\widehat I_\alpha,\widehat\HH^j(s)]=0
\end{equation}
for any $\alpha\in\g$. Hence in these Examples we have a homomorphism of Lie algebras $\g\to A_h$, $\alpha\mapsto i\widehat I_\alpha/h$.

In these Examples, it is easy to show that the $G$-action on $\Delta_h$ given by (\ref{canon}) preserves 
the operation $*_h$ and the elements $\widehat\HH^j(s)$, and generates the $\g$-action on $A_h$ given by commutator with $i\widehat I_\alpha/h$, 
so that $A_h$ is a $G$-equivariant associative algebra.

\section{Examples: free fields}

In this Section we study quantizations $A_h$ for free scalar field in the Minkowski spacetime and 
for the Dirichlet principle.

\subsection{Free scalar field in the Minkowski spacetime}
\label{free_scalar_field}
It is given by (\ref{scalar_field},  \ref{nonparam_HJ_k}, \ref{HJ_scalar_field}--\ref{D^mu^nu}) with
\begin{equation}
\label{free_m}
V(\varphi)=\frac{m^2}2 \varphi^2.
\end{equation} 
The number $m$ is called the mass.

The Euler--Lagrange equation is the Klein--Gordon equation
\begin{equation}
\label{Klein_Gordon}
\frac{\partial^2\varphi}{(\partial x^0)^2}-\sum_{j=1}^n\frac{\partial^2\varphi}{(\partial x^j)^2}+m^2\varphi=0.
\end{equation}

According to the formalism of \S\ref{classical}, the Poisson algebra $A_0$ of classical observables, i.~e. functionals of a solution $\varphi(x)$ of the Klein--Gordon equation,
is identified with the Poisson algebra of functionals $\Phi(x^j(\cdot)$, $\varphi(\cdot)$, $\pi(\cdot))$ of Cauchy data $x^j(s)$, $\varphi(s)$, $\pi(s)$, 
satisfying equations (\ref{equation_for_nonparametric_observables}), where $x^j=x^j(s)$ 
is a space-like surface in the spacetime $\R^{n+1}$.   

It is easy to see that quantization $A_h$ (\ref{A_h}--\ref{Schrod}) is identified with the algebra of solutions $\Phi^{(h)}(x^j(\cdot),\varphi(\cdot),\pi(\cdot))$ of generalized Heisenberg equation (\ref{gen_Heisenberg})
with values in the Weyl algebra $\Delta_h^0$ of functionals $\Phi(\varphi(\cdot),\pi(\cdot))$ with the Moyal product, cf.~(\ref{Moyal}), and with $\widehat H_j(s)=H_j(s)$.

The generalized Heisenberg equation is, in general, not well defined. However, for free field (\ref{free_m}) Hamiltonian densities (\ref{Ham_density}) are quadratic forms of $(\varphi(\cdot),\pi(\cdot))$, and 
the generalized Heisenberg equation is well defined. Moreover, commutator in (\ref{gen_Heisenberg}) coincides with the Poisson bracket multiplied by $-ih$. Therefore the algebra $A_h$ coincides with $A_0$ as a vector space.
As an algebra, $A_h$ is the Weyl algebra of the symplectic vector space of solutions $\varphi(x)$ of Klein--Gordon equation (\ref{Klein_Gordon}).

This algebra coincides with the algebra of observables in free algebraic traditional QFT [\ref{TV}]. For two dimensional spacetime, i.~e. for $n=1$, generalized Heisenberg equation
(\ref{gen_Heisenberg}) corresponds to a generalized Schr\"odinger equation (\ref{gen_Schrod}) with values in the Fock space $W_0$ (the Tomonaga--Schwinger equation). However, for $n>1$ the generalized 
Schr\"odinger equation is absent, see [\ref{TV}].

\subsection{The Dirichlet principle} 
It is given by (\ref{Dirichlet}, \ref{nonparam_HJ_k}, \ref{HJ_Dirichlet}).

The Euler--Lagrange equation is the Laplace equation
\begin{equation}
\frac{\partial^2\varphi}{(\partial x^0)^2}+\frac{\partial^2\varphi}{(\partial x^1)^2}=0.
\end{equation}
We shall consider harmonic functions, i.~e. solutions $\varphi(x)$ of the Laplace equation, on the plane $x=(x^0,x^1)$ without the origin $(0,0)$.

According to the formalism of \S\ref{classical}, the Poisson algebra $A_0$ of classical observables, i.~e. functionals of a harmonic function, is identified with the Poisson algebra of functionals 
$\Phi(x^0(\cdot),x^1(\cdot),\varphi(\cdot),\pi(\cdot))$
of functions $x^0(s),x^1(s),\varphi(s),\pi(s)$ of a variable $s\in\R/\Z$, i.~e. of a real variable $s$ periodic with period 1, satisfying equations (\ref{equation_for_nonparametric_observables}), where $(x^0(s),x^1(s))$ is a
smooth embedding of the circle $\R/\Z$ into the plane without origin $x\ne0$ surrounding the origin counterclockwise. 

As in \S\ref{free_scalar_field}, quantization $A_h$ (\ref{A_h}--\ref{Schrod}) is identified with the algebra of solutions $\Phi^{(h)}(x^0(\cdot),x^1(\cdot),\varphi(\cdot),\pi(\cdot))$ of generalized Heisenberg equation (\ref{gen_Heisenberg})
with values in the Weyl algebra $\Delta_h^0$ of functionals $\Phi(\varphi(\cdot),\pi(\cdot))$ with the Moyal product, and with $\widehat H_j(s)=H_j(s)$, $j=0,1$.

The generalized Heisenberg equation is well defined, because the Hamiltonian densities $H_0(s)$, $H_1(s)$ are quadratic forms in $(\varphi(\cdot),\pi(\cdot))$.
Moreover, commutator in (\ref{gen_Heisenberg}) coincides with
the Poisson bracket multiplied by $-ih$. Therefore the algebra $A_h$ coincides with $A_0$ as a vector space. As an algebra, $A_h$ is the Weyl algebra of the symplectic vector space of harmonic functions $\varphi(x)$.

This picture is simplified if we pass to complex variables. Put
\begin{equation}
\HH(s)=\frac12(\HH^1(s)+i\HH^0(s)),\ \ \overline\HH(s)=\frac12(\HH^1(s)-i\HH^0(s)),
\end{equation}
and
\begin{equation}
\label{complex_var} 
\begin{aligned}
&z(s)=x^0(s)+ix^1(s),\ \ \overline z(s)=x^0(s)-ix^1(s),\\
&H(s)=\frac12(H_0(s)-iH_1(s)),\ \ \overline H(s)=\frac12(H_0(s)+iH_1(s)),\\
&\rho(s)=\sqrt{\frac i2}(\pi(s)-i\varphi_s(s)),\ \ \overline\rho(s)=\sqrt{\frac{-i}2}(\pi(s)+i\varphi_s(s)).
\end{aligned}
\end{equation}
Then we have
\begin{equation}
\HH(s)=-z_s(s)H(s)+\frac12\rho(s)^2,\ \ \overline\HH(s)=-\overline z_s(s)\overline H(s)+\frac12\overline\rho(s)^2,
\end{equation}
and
\begin{equation}
\begin{aligned}
&[z(s),H(s')]=[\overline z(s),\overline H(s')]=-ih\delta(s-s'),\\
&[\rho(s),\rho(s')]=[\overline\rho(s),\overline\rho(s')]=ih\delta'(s-s'),
\end{aligned}
\end{equation}
and all the rest commutators of variables in (\ref{complex_var}) vanish. Hence a computation gives
\begin{equation}
\begin{aligned}
&[\HH(s),\HH(s')]=ih\delta'(s-s')(\HH(s)+\HH(s')),\\
&[\overline\HH(s),\overline\HH(s')]=ih\delta'(s-s')(\overline\HH(s)+\overline\HH(s')),\\
&[\HH(s),\overline\HH(s')]=0,
\end{aligned}
\end{equation}
i.~e. $\HH(s)$ and $\overline\HH(s)$ span two commuting copies of the complexified Lie algebra of vector fields on the circle.

This implies that the algebra $A_h$ is the (topological) tensor product of mutually commuting subalgebras $A^c_h$ and $\overline{A^c_h}$, where $A^c_h\subset A_h$ is the algebra of analytical functionals 
$\Phi(z(\cdot),\rho(\cdot))$ satisfying the Heisenberg equation
\begin{equation}
\label{complex_Heis}
ihz_s(s)\frac{\delta\Phi}{\delta z(s)}=\left[\frac12\rho(s)^2,\Phi\right],
\end{equation}
and $\overline{A^c_h}$ is the similar algebra in complex conjugate variables.

If we identify the algebra $A_h$ with the Weyl algebra of functionals $\Phi(\varphi(\cdot))$ on the symplectic vector space of harmonic functions $\varphi(x)$, $x\ne0$, then the subalgebra $A^c_h\subset A_h$ is identified with the 
Weyl algebra of functionals $\Phi(\varphi_z(\cdot))$ on the symplectic vector space of holomorphic functions
\begin{equation}
\varphi_z(z)=\frac{\partial\varphi}{\partial z}\left(x^0,x^1\right)=\frac12\left(\frac{\partial\varphi}{\partial x^0}-i\frac{\partial\varphi}{\partial x^1}\right)\left(x^0,x^1\right),\ \ z=x^0+ix^1,
\end{equation}
on the complex plane $z\ne0$ without origin. A functional $\Phi(\varphi_z(\cdot))\in A^c_h$ is identified with a functional $\Phi(z(\cdot),\rho(\cdot))$ satisfying Heisenberg equation 
(\ref{complex_Heis}) by means of the formula 
\begin{equation}
\rho(s)=\sqrt{\frac 2i}\varphi_z(z(s))z_s(s).
\end{equation}

The algebra $\overline{A^c_h}$ has similar description in terms of antiholomorphic function
\begin{equation}
\varphi_{\overline z}(\overline z)=\frac{\partial\varphi}{\partial\overline z}\left(x^0,x^1\right)=\frac12\left(\frac{\partial\varphi}{\partial x^0}+i\frac{\partial\varphi}{\partial x^1}\right)\left(x^0,x^1\right),\ \ \overline z=x^0-ix^1.
\end{equation}

Heisenberg equation (\ref{complex_Heis}) corresponds to the Schr\"odinger equation
\begin{equation}
ihz_s(s)\frac{\delta\Psi}{\delta z(s)}=\frac12\rho(s)^2\Psi
\end{equation}
for a functional $\Psi(z(\cdot))$ with values in the Fock space $W_0$. This yields a projective action of the Lie algebra of vector fields on a circle on the Fock space $W_0$, called 
free field representation in two dimensional conformal field theory [\ref{BPZ}, \ref{FF}, \ref{Fr}].

\section*{Conclusion}

We have constructed a mathematical theory of interacting (bosonic) quantum fields. 
The author is not a physicist, and finding out the physical sense of our theory is a question to physicists.

In traditional QFT the algebra of observables is a $*$-algebra, i.~e. an algebra with involution, which acts on a Hilbert space of states.
It seems that, in general, it is impossible to define the involution for algebras constructed in the present paper. Indeed, if we consider the time variable and the coordinate variables in the Schr\"odinger equation
on equal rights, then the definition of the Hilbert space of quantum mechanical states and of the involution in the algebra of observables becomes artificial.

However, we expect that in low spacetime dimensions, known rigorous models of traditional QFT in the Hilbert space of states, like two dimensional conformal 
field theory, amount to the theory constructed in the present paper. The situation here 
is a generalization of the situation with theory of the wave equation (equation (\ref{Klein_Gordon}) with $m=0$). 
In two dimensions the wave equation can be solved in the Hilbert space of usual functions, due to separation of variables. However, for higher
dimensions the Hilbert space is insufficient, and one has to use distributions. Similarly, in known low dimensional rigorous QFT's one has a Hilbert space of states, but 
in general QFT's the Hilbert space is absent. Instead, one has an associative algebra of observables. 

A very interesting question is whether for general QFT's there exists the analog of distributions,
i.~e. the space of generalized states, forming a representation of the algebra of observables. As far as we can see now, such a space does not exist even for free scalar field in spacetimes of dimension greater than two.


\begin{thebibliography}{99}
\bibitem{}\label{BPZ} A. A. Belavin, A. M. Polyakov, and A. B. Zamolodchikov, Infinite conformal symmetry in two-dimensional quantum field theory, Nucl. Phys. B241 (1984), 333--380.
\bibitem{}\label{BS} N. N. Bogoliubov and D. V. Shirkov, Introduction to the theory of quantized fields, Interscience, New York, 1959.
\bibitem{}\label{D} P. A. M. Dirac, General theory of relativity, Princeton University Press, 1975.
\bibitem{}\label{FF} B. L. Feigin and D. B. Fuks, Representations of the Virasoro algebra, in: Representations of Infinite Dimensional Lie Groups and Lie Algebras, Gordon and Breach, 1989.
\bibitem{}\label{Fr} E. Frenkel, Affine Kac--Moody algebras at the critical level and quantum Drinfeld--Sokolov reduction, Ph. D. Thesis, Harvard, 1991.
\bibitem{}\label{IZ} C. Itzykson and J. B. Zuber, Quantum field theory, McGraw Hill, 1980. 
\bibitem{}\label{MS} V. P. Maslov and O. Yu. Shvedov, Method of complex germ in the many-particle problem and in quantum field theory, Editorial URSS, Moscow, 
2000 (in Russian).
\bibitem{}\label{St07} A. Stoyanovsky, Introduction to the mathematical principles of quantum field theory, Editorial URSS, Moscow, 1st edition, 2007;
2nd edition, 2015 (in Russian).
\bibitem{}\label{TV} C. G. Torre and M. Varadarajan, Functional evolution of free quantum fields, arXiv:hep-th/9811222, Class. Quant. Grav. 16 (1999) 2651--2668.
\end{thebibliography}
\end{document}